\documentclass[useAMS,usenatbib,usegraphicx]{mn2e}
\usepackage{epsf,rotating,url}
\usepackage{subfigure,amssymb,amsfonts,amstext,amsgen,amsopn,amsxtra,indentfirst,lscape,times,rotating}
\usepackage{longtable}
\usepackage{graphics}
\usepackage{keyval}
\usepackage{trig}
\usepackage{dcolumn}
\usepackage{bm}
\def\beq{\begin{equation}}
\def\eeq{\end{equation}}
\def\bey{\begin{eqnarray}}
\def\eey{\end{eqnarray}}
\def\msun{M_\odot}

\def\kms{\, {\rm km \, s}^{-1} }

\def\prd{Phys. Rev. D}
\def\mnras{MNRAS}
\def\apj{ApJ}

\def\apjs{ApJ}
\def\apjl{ApJ}

\def\physrep{Physics Reports}
\def\aap{A \& A}

\def\lcdm{{\Lambda}CDM}

\def\aap{Astron. Astrophys.}

\title{The abundance of galaxy clusters in MOND: Cosmological simulations with massive neutrinos}

\author[G. W. Angus, A. Diaferio]{G. W. Angus$^{1,2,3}$\thanks{E-mail: angus.gz@gmail.com} and Antonaldo Diaferio$^{2,3,4}$ \\ 
$^{1}$Astrophysics, Cosmology \& Gravity Centre, University of Cape Town, Private Bag X3, Rondebosch, 7700, South Africa \\
$^{2}$Dipartimento di Fisica Generale ``Amedeo Avogadro", Universit\`a degli studi di Torino, Via P. Giuria 1, I-10125, Torino, Italy \\
$^{3}$Istituto Nazionale di Fisica Nucleare (INFN), Sezione di Torino, Torino, Italy\\
$^{4}$Harvard-Smithsonian Center for Astrophysics, 60 Garden Street, Cambridge, MA 02138, USA \\}
\begin{document}

\date{\today}
\maketitle
\begin{abstract}
We present a new Particle-Mesh cosmological N-body code for accurately solving the modified Poisson equation of the Quasi Linear formulation of MOND. We generate initial conditions for the Angus (2009) cosmological model, which is identical to $\lcdm$ except that the cold dark matter is switched for a single species of thermal sterile neutrinos. We set the initial conditions at $z=250$ for a $(512~Mpc/h)^3$ box with $256^3$ particles and we evolve them down to $z=0$. We clearly demonstrate the necessity of MOND for developing the large scale structure in a hot dark matter cosmology and contradict the naive expectation that MOND cannot form galaxy clusters. We find that the correct order of magnitude of X-ray clusters (with $T_X > 4.5~keV$) can be formed, but that we overpredict the number of very rich clusters and seriously underpredict the number of lower mass clusters. The latter is a shortcoming of the resolution of our simulations, whereas we suggest that the over production of very rich clusters might be prevented by
incorporating a MOND acceleration constant that varies with redshift and an expansion history that cannot be described by the usual Friedmann models. We present evidence that suggests the density profiles of our simulated clusters are compatible with those of observed X-ray clusters in MOND. It remains to be seen if the low mass end of the cluster mass function can be reproduced and if the high densities of dark matter in the central $20~kpc$ of groups and clusters of galaxies, measured in the MOND framework, can be achieved. As a last test, we computed the relative velocity between pairs of halos within $10~Mpc$ and find that pairs with velocities larger than $3000~\kms$, like the bullet cluster, can form without difficulty.
\end{abstract}


\section{Introduction}
\protect\label{sec:intr}
Cosmological simulations and other calculations of the formation of large scale structure in the Modified Newtonian Dynamics (MOND) of \cite{milgrom83a} have a substantial literature, despite the absence of an agreed upon cosmological model. Early work by \cite{sanders98} crucially demonstrated that the representation of the expansion of the Universe by a scale factor, derived from the Friedmann-Robertson-Walker (FRW) metric, is justified in MOND because expansion will indeed be uniform, contrary to earlier claims by \cite{felten84}. In addition to this, the results of Big Bang Nucleosynthesis (assuming the number of neutrinos is not modified) were shown to be unaltered by MOND, and the natural result that structure formation proceeds more rapidly in MOND was verified. 

\cite{sanders01} continued this theme by modifying the CMBFAST code of \cite{seljak96}  for a two field MOND-like Lagrangian based theory and showed that, by and large, the observed distribution of galaxies (the power spectrum at the time) appeared to be reproduced in a low density, baryon only, vacuum energy dominated model. Unfortunately for that model, today's precision cosmology contradicts it at high significance, but nevertheless, a solid basis was set which proved matter over-densities, which evolve ponderously in Newton's gravity, can evolve with great vigour in Milgrom's modified dynamics (see his Fig~2).

Following these pivotal works, \cite{nusser02} made the first foray into MOND cosmological simulations by modifying a particle-mesh (PM) code to include an approximation to the MOND equation and then performed a series of cosmological simulations. Unfortunately, the simulations were run without any vacuum energy, but generally showed the alarming trend for MOND to produce too much structure on the scales $k=0.1-1.0~k/Mpc$. This conclusion prevailed even in an open Universe with $\Omega_m$=0.03 and the acceleration constant of MOND being reduced to $1/12$ of its typical value. One thing that is uncertain is whether or not the initial conditions were self-consistent for a distribution of baryons in a low density Universe. For instance, the power spectrum at redshift $z=100$ for $k>0.5h/Mpc$ should be less than $10^{-10}k^3/Mpc^3$, in contrast to a cold dark matter dominated Universe ($\Omega_m=1.0$) for which $P(k)>10^{-2}$ for all $k<3.0h/Mpc$.

\cite{knebe04} incorporated the algebraic MOND equation into their Multi Level Adaptive Particle Mesh code (called MLAPM) and, using it, made some similar comparisons to the $\lcdm$ model with MOND as \cite{nusser02}. 

In a seminal piece of work, \cite{skordis06} linearised the equations of motion from \cite{bekenstein04} and incorporated them into CMBFAST to derive the linear evolution of the formation of structure in that theory. They showed that, with the addition of 3 species of $2.75~eV$ active neutrinos (and a substantially increased MOND acceleration constant), the galaxy power spectrum, as measured by the SDSS survey, could be matched by the evolved power spectrum from the code. Whether galaxies or clusters would go on and form the correct distribution is unknown, but highly doubtful, because shortly thereafter \cite{dodlig} performed a similar calculation and showed that in the framework of TeVeS, there was no way to form the correct distribution of cosmic structure without non-baryonic dark matter.

Leaving aside relativistic counterparts of MOND, a crucial development on the MLAPM code was made by \cite{llinares08} and further used in \cite{llinares09} because, instead of ``merely" solving the algebraic MOND equation, they solved the momentum and energy conserving modified Poisson equation of the \cite{bekenstein84} theory using the same technique as originally outlined by \cite{brada99a} and further used by \cite{tcevol} for galaxy simulations.

Although incredibly important stepping stones, the obvious flaw with all these previous works is that, although they solve some variant on the MOND equation in their simulations, they start from arbitrary initial conditions. None were derivatives of models that were consistent with observations of the anisotropies in the cosmic microwave background (CMB) nor the presence of dark matter in clusters of galaxies. This no longer has to be the case because \cite{angus09} has proposed a model that can allow MOND to satisfy many cosmological constraints.

The model currently employs a simple ansatz - that whichever relativistic version of MOND is correct, the FRW metric is still the correct description of the expansion history of the Universe, which is a perfectly reasonable assumption expected from relativistic analogues of MOND (\citealt{feix10,clifton10}, see also \citealt{skordis11}). Given that typical accelerations at high $z$ are of the order of the MOND acceleration constant, this is a justified simplification at those redshifts. Even were it not, there is no reason to believe that the MOND acceleration constant, which is suggested to be linked to the presence of dark energy, need be constant with redshift. There is, however, the very logical possibility that the expansion history cannot be described by a standard Friedmann model and this is a topic that is addressed further in \S\ref{sec:massfunc}.

The idea behind the model is that MOND has been argued to be consistent with the dynamics of all galactic and sub-galactic dynamics, including the dwarf spheroidal galaxies surrounding the Milky Way (\citealt{angus08}) and the recently analysed THINGS rotation curves (\citealt{gentile11}), without the need for cold dark matter. Although there are still contentious issues surrounding these systems, the balance of evidence suggests MOND can at least do as good a job of fitting their dynamics as a cold dark matter halo which has several extra free parameters. When this is added to the evidence from the baryonic Tully-Fisher relation (\citealt{mcgaugh11}) it provides a strong case that MOND is a good description of dynamics on these sub megaparsec scales.

However, when MOND is extended to cosmological data sets, it seems to comprehensively fail. Apart from the need for dark energy, it requires some form of dark matter to fix the dynamics of clusters of galaxies and expansion history, as well as the CMB. It is therefore obvious that if MOND is to be preserved as a good model of galactic dynamics, and at the same time have some measure of success in fitting cosmological data, it must do so in cooperation with dark matter. The constraint on this dark matter is that it must be hot enough to free stream from galaxies such that it does not destroy the good fits to galactic rotation curves. \cite{angus09} showed that several cosmological constraints can be satisfied using an $11~eV$ sterile neutrino that is fully thermalised in the early Universe.

\begin{figure}
\includegraphics[angle=0,width=8.0cm]{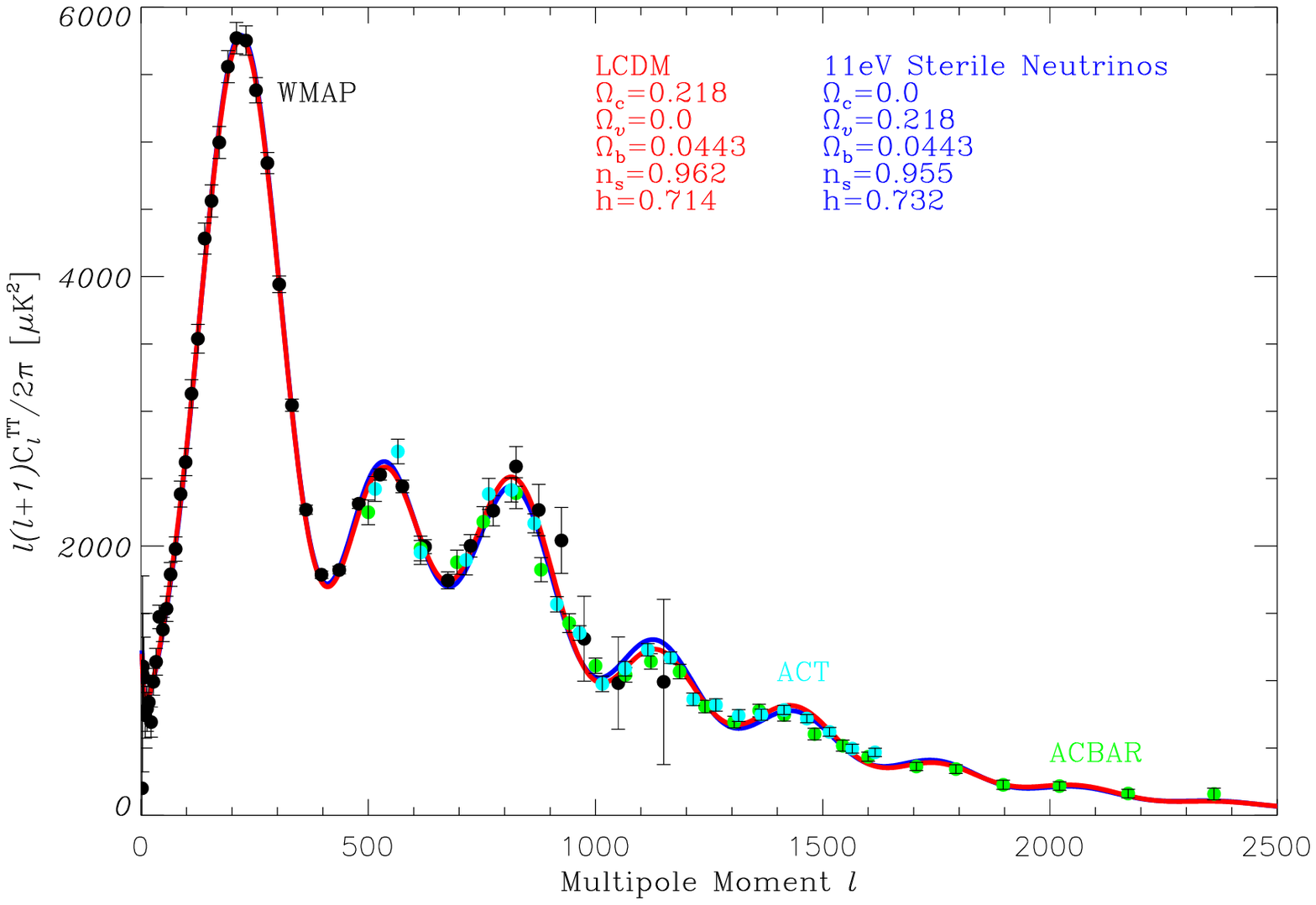}
\caption{Shows the CMB angular power spectrum for our cosmological model (blue line), compared with the $\lcdm$ model (red line). The data points come from WMAP year 7 (black), ACT (turquoise) and ACBAR (green).  }
\label{fig:cmb}
\end{figure}

For example, it allows an excellent fit to the angular power spectrum of the CMB (reproduced in Fig~\ref{fig:cmb}) by virtue of $\Omega_{\nu_s}$ for an $11~eV$ sterile neutrino being almost identical to $\Omega_{cdm}$ of the $\lcdm$ model. To complement this, \cite{afd} demonstrated that the dark matter density profiles found in MOND of the sample of groups and clusters of galaxies studied by \cite{afb} were consistent with equilibrium configurations of $11~eV$ sterile neutrinos. What this means is that the velocity dispersions of the sterile neutrinos required for equilibrium of the density profiles form a phase space density that does not need to exceed the Tremaine-Gunn limit.

Although the phase space densities of sterile neutrinos are compatible with static clusters of galaxies, the obvious next step is to investigate whether the model can form the observed distribution and internal structure of clusters of galaxies in the finite timescale allowed by the age of the Universe. Equally as essential, given that our model uses hot dark matter, is whether the matter power spectrum is matched on the smaller scales measured by the Lyman alpha forest (e.g. \citealt{tegmark04}) at high redshift.

Clearly, the only way to test these issues is with cosmological numerical simulations, like those of \cite{nusser02,knebe04,llinares08} except that the initial conditions will correspond, as near as possible, to the cosmological model of \cite{angus09} - hereafter Angus09. In this paper we developed a single coarse grid cosmological particle-mesh code and then adapted it to solve the modified Poisson equation of Milgrom's recently proposed QUasi-linear MOND (QUMOND, \citealt{milgrom10}). QUMOND is a new incarnation of MOND that is more convenient to handle due to the fact that the theory only requires solutions of a linear differential equation and one non-linear algebraic step. It can be derived from an action, thus the conservation laws are adhered to.

Using this QUMOND code, we run a single cosmological simulation with as high resolution
as possible and compare it with the observed distribution of mass on scales of galaxy
clusters and beyond. Unfortunately, we do not have the resolution to investigate
smaller scales that are dominated by baryonic physics.

\section{QUMOND}
\protect\label{sec:code}

The Aquadratic Lagrangian theory of \cite{bekenstein84} produces a modified Poisson equation which must be solved numerically for arbitrary geometries. Likewise, QUMOND requires solution of a modified Poisson equation, but one that is slightly easier to implement. Specifically, the ordinary Poisson equation for cosmological simulations

\beq
\protect\label{eqn:qumond1}
\nabla^2\Phi_N=4\pi G (\rho - \bar{\rho})/a
\eeq
is solved to give the Newtonian potential, $\Phi_N$, at scale factor $a$, from the ordinary matter density $\rho$ which includes baryons and neutrinos. This would also include cold dark matter if there was any in our model. The QUMOND potential, $\Phi$, is found from the Newtonian potential as follows
\beq
\protect\label{eqn:qumond2}
\nabla^2\Phi=\nabla . \left[ \nu(y) \nabla\Phi_N \right],
\eeq
where $\nu(y)=0.5+0.5\sqrt{1+4/y}$ and $y=\nabla\Phi_N/a_o a$, with $a_o$ being the MOND acceleration constant chosen here to be $3.6~(\kms)^2pc^{-1}$.

Essentially, we solve the Poisson equation once, using the ordinary matter density as the source, as would happen in a regular N-body code, to find the Newtonian potential (Eq~\ref{eqn:qumond1}) - incorporating periodic boundary conditions. Following this, we take derivatives of the Newtonian potential and generate a new source density (the right hand side of Eq~\ref{eqn:qumond2}), usually referred to as the ordinary matter plus phantom matter density associated with the QUMOND field. Then we solve the Poisson equation a second time to find the QUMOND potential, $\Phi$. From this, we take the gradient of the QUMOND potential to find the gravitational field at each particle's location and move our particles accordingly.

The density assignment follows the standard cloud-in-cell technique and from this both Poisson equations (Eqs~\ref{eqn:qumond1} and \ref{eqn:qumond2}) are solved using Multigrid methods. The densities found from backwards derivation of the potentials gives the source density back with impeccable precision at all redshifts and locations in the grid. Next comes the important part where the QUMOND source density is found.

Specifically, what happens is the following: assume we want the QUMOND source at cell ($i$,$j$,$k$) of a Cartesian grid ($x$,$y$,$z$), then we need to define the gravity at various points surrounding it. If we use unit length grid cells then
\bey
\nonumber g_{x_2} = \phi_{i+1,j,k} - \phi_{i,j,k}\\
\nonumber g_{x_1} = \phi_{i,j,k} - \phi_{i-1,j,k}\\
\nonumber g_{y_2} = \phi_{i,j+1,k} - \phi_{i,j,k}\\
\nonumber g_{y_1} = \phi_{i,j,k} - \phi_{i,j-1,k}\\
\nonumber g_{z_2} = \phi_{i,j,k+1} - \phi_{i,j,k}\\
g_{z_1} = \phi_{i,j,k} - \phi_{i,j,k-1}
\eey
These are the values of the gravitational field at half a cell from ($i$,$j$,$k$) in the three orthogonal directions. Similarly, for these six points we must find the value of the $\nu$ function. Surrounding the point $x_2$ we use
\bey
 \omega_{x_2}&=&\phi_{i+1,j,k} - \phi_{i,j,k}\\
\nonumber 4\omega_{y_2}&=&\phi_{i+1,j+1,k} + \phi_{i,j+1,k} - \left(\phi_{i+1,j-1,k} + \phi_{i,j-1,k} \right)\\
\nonumber 4\omega_{z_2}&=&\phi_{i+1,j,k+1} + \phi_{i,j,k+1} - \left(\phi_{i+1,j,k-1} + \phi_{i,j,k-1} \right)
\nonumber
\eey
and surrounding $x_1$
\bey
 \omega_{x_1}&=&\phi_{i,j,k} - \phi_{i-1,j,k}\\
\nonumber 4\omega_{y_1}&=&\phi_{i,j+1,k} + \phi_{i-1,j+1,k} - \left(\phi_{i,j-1,k} + \phi_{i-1,j-1,k} \right)\\
\nonumber 4\omega_{z_1}&=&\phi_{i,j,k+1} + \phi_{i-1,j,k+1} - \left(\phi_{i,j,k-1} + \phi_{i-1,j,k-1} \right)
\nonumber
\eey
which gives
\bey
\nonumber \kappa_{x_2}=(a_oa)^{-1}\sqrt{\omega_{x_2}^2+\omega_{y_2}^2+\omega_{z_2}^2}\\
 \kappa_{x_1}=(a_oa)^{-1}\sqrt{\omega_{x_1}^2+\omega_{y_1}^2+\omega_{z_1}^2}.
\eey
These are accompanied by $\kappa_{y_2}$, $\kappa_{y_1}$, $\kappa_{z_2}$ and $\kappa_{z_1}$ found in a similar way. From this we must find 
\bey
\nonumber \nu_{x_2}=\nu(\kappa_{x_2})\\
\nu_{x_1}=\nu(\kappa_{x_1})
\eey
and $\nu_{y_2}$, $\nu_{y_1}$, $\nu_{z_2}$, $\nu_{z_1}$. This finally leaves us with the QUMOND source density in cell ($i$,$j$,$k$) given by $\nu_{x_2}g_{x_2} - \nu_{x_1}g_{x_1} + \nu_{y_2}g_{y_2} - \nu_{y_1}g_{y_1}+ \nu_{z_2}g_{z_2}- \nu_{z_1}g_{z_1}$. A good visualisation of the geometry can be found in \cite{tcevol} or \cite{llinares08}.

\section{Initial Conditions}
\protect\label{sec:ics}
The cosmological model we use employs ($\Omega_b$, $\Omega_{\nu_s}$, $\Omega_{\Lambda}$, $h$, $n_s$)=(0.0443, 0.218, 0.7377, 0.732, 0.955), which means that one thermal sterile neutrino is required with a mass near to 11~$eV$. We plot in Fig~\ref{fig:cmb} a comparison between our model's fit to the CMB and the $\lcdm$ model's. We note that our model requires four effective neutrino species (3 active and 1 sterile), and as such the $^4He$ fraction expected from big bang nucleosynthesis is 0.261, which is roughly $1-\sigma$ larger than the measured value (see e.g. \citealt{mangano11}). The other relic abundances are not negatively altered and the value for $^7Li$ is sadly not improved.

We advocate starting simulations before $z \sim 100$ where MOND effects are minimal, and this means producing initial conditions at these redshifts. As mentioned previously, in the absence of a theoretical underpinning of MOND, and given that there is no consensus on a relativistic theory, we must produce the initial conditions under the ansatz that gravity does not deviate from general relativity while $z>100$. This allows us to use the COSMICS package of \cite{bert} without modification.

\begin{figure}
\def\subfigtopskip{0pt} 
\def\subfigbottomskip{4pt}
\def\subfigcapskip{1pt}
\centering
\begin{tabular}{c}
\subfigure{
\includegraphics[angle=0,width=8.0cm]{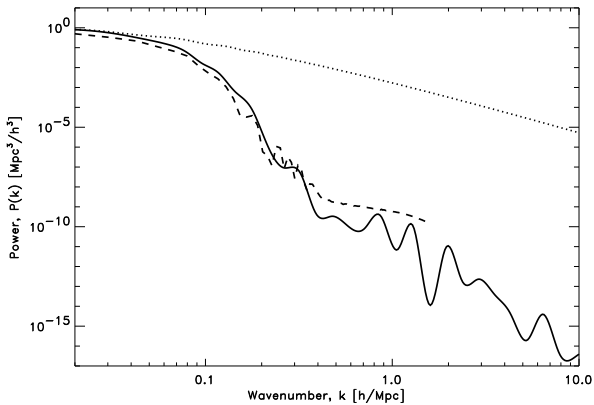}
}\\
\end{tabular}

\caption{Shows the power spectrum of our N-body realisation (dashed) at $z=254.1$, the expected power spectrum from the CAMB package (solid) for that model as well as the power spectrum for the $\lcdm$ model (dotted line).}
\label{fig:mps200}
\end{figure}

The initial power spectrum of the Angus09 model was computed accurately at $z=254.1$ using the CAMB package of \cite{lewis} which can be compared (see Fig~\ref{fig:mps200}) to the power spectrum of our N-body initial conditions which are generated using the COSMICS package of \cite{bert}. Both codes accurately take care of the neutrino distribution function, but of course the COSMICS package can only represent the power spectrum discretely by distributing particles (in our case $256^3$) on a $257^3$ grid of length $512~Mpc/h$. Each particle weighs $7.8\times10^{11}\msun$. 

The N-body realisation has its amplitude normalised by the CMB quadrupole, $Q_{rms-PS}=T_0\left({5C_2 \over 4\pi}\right)^{1 \over 2}$, where $T_0=2.725~K$ is the current temperature of the CMB and $C_2$ is the $l=2$ component of the angular power spectrum. This is enforced because we cannot use linear theory in MOND to estimate $\sigma_8$ at $z=0$. According to \cite{bennett11}, the 1-$\sigma$ range of $Q_{rms-PS}=17-45~\mu K$, so to be highly conservative, we use $18~\mu K$. For comparison, in a $\lcdm$ universe, $Q_{rms-PS}=18~\mu K$ corresponds to $\sigma_8=0.68$ and the typically used $\sigma_8=(0.75, 0.8, 0.9)$ correspond to $Q_{rms-PS}=(17.5, 18.6, 21.0)~\mu K$ respectively. The dependence of the formation of structure on this parameter in the Angus09 model will be tested in a forthcoming paper.

Perhaps more significant than this, one can see in Fig~\ref{fig:mps200} that the true power spectrum from CAMB is not perfectly reproduced by the simulation initial conditions. In particular, the baryonic acoustic peaks and the exact turn off due to the neutrino free streaming are not replicated exactly. Nevertheless, when the theoretically expected power declines due to the neutrino free streaming (around $k = 0.1~h/Mpc$) the simulated power drops. Another obvious feature is that the simulated power at $k > 0.3~h/Mpc$ is somewhat greater than the correct power, however, the power is still incredibly small. For instance, for the Angus09 model $P(k=0.7~h/Mpc)=10^{-9}Mpc^3/h^3$, which can be compared to a $\lcdm$ model (where our $\Omega_{\nu_s}$ is directly swapped for $\Omega_{cdm}$) which has power at $k=0.7~h/Mpc$ of $10^{-3}Mpc^3/h^3$ - some six orders of magnitude larger.

\section{The importance of MOND for structure growth}
\protect\label{sec:ps}
With standard $\lcdm$ simulations, we know that from $z=250$ to 0 the perturbations need only grow by a factor of around $10^5$ or so at any wavenumber. In contrast, we know that to be viable models, the sterile neutrino perturbations must increase by 15 orders of magnitude at large wavenumbers.

For our test case we took our initial conditions discussed in \S\ref{sec:ics} and ran three simulations: one using the code described in \S\ref{sec:code} and another using only the Newtonian potential i.e. with MOND turned off. We also ran a simulation using the MLAPM code of \cite{knebe04} with the simple $\mu$ function (\citealt{fb05,famaey06}). To be clear, this code does not properly solve the Bekenstein-Milgrom modified Poisson equation (it ignores the curl field), but rather only the algebraic relation $\nabla \Phi = \mu(\nabla \Phi)\nabla \Phi_N$, where $\mu$ is also an interpolating function, similar to the $\nu$ used in this paper. It is these three simulations that we discuss in the remainder of the paper.

\subsection{Power Spectrum}
\begin{figure}
\def\subfigtopskip{0pt} 
\def\subfigbottomskip{4pt}
\def\subfigcapskip{1pt}
\centering
\begin{tabular}{c}
\subfigure{
\includegraphics[angle=0,width=8.0cm]{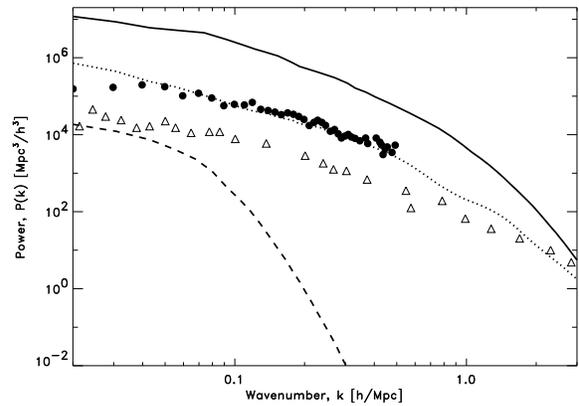}
}\\
\end{tabular}

\caption{Shows the measured power spectrum of galaxy clusters at $z=0$ from the REFLEX II survey (filled circles) and for reference the galaxy power spectrum (open triangles) from Tegmark et al. (2004). We also plot the power spectra of particles from our three simulations: QUMOND using our code (solid), algebraic MOND using MLAPM (dotted) and without MOND (dashed).}
\label{fig:mps0}
\end{figure}

In Fig~\ref{fig:mps0} we plot the $z=0$ power spectrum of the particles in the three simulations described above along with the measured power spectrum of clusters of galaxies from the REFLEX II survey (\citealt{reflex2}) and also the power spectrum of galaxies (\citealt{tegmark04}). The final power spectrum of the simulation without MOND falls drastically short of the measured power spectrum of clusters, and other probes, at all wavenumbers. This is the reason why such models with hot dark matter and untampered gravity are correctly ignored in the literature.

This power spectrum without MOND can be compared to the power spectra found with our QUMOND code and the MLAPM simulation where we find that all wavenumbers probed increase in power to amplitudes far larger than the power spectrum of galaxies, and, in the case of the QUMOND simulation, far larger than the clusters of galaxies. This is potentially problematic, but there exist parameters that can be feasibly altered to slow the growth of structure - for instance the acceleration constant of MOND decreasing with increasing redshift, which is discussed in \S\ref{sec:massfunc}.  The serendipitous match between the MLAPM simulation and the power spectrum of clusters may suggest that the gravity missing due to the non-inclusion of the curl-field is enough to bring the QUMOND simulation into agreement. Alternatively, the disparity may be a result of not comparing like for like. For instance, in $\lcdm$, the growth of structure is hierarchical and similar on all scales because cold dark matter has, by definition, power on all scales. Conversely, both MOND and sterile neutrinos have a scale dependence.

MOND is acceleration dependent and gravity is therefore a product of environment, and the sterile neutrinos have a length scale below which power is exponentially suppressed. This length scale is larger than the diameter of galaxies and thus sterile neutrinos have merely a shepherding influence on the baryons that build galaxies. For example, the sterile neutrinos collapse (together with the baryons) into filamentary structures, with clusters of galaxies forming at the nodes of these filaments. At this point, galaxies can only form if a perturbation from tidal torques or some other shock to the primordial gas can create a perturbation which can grow under MOND. Obviously this cannot occur efficiently at the nodes where the galaxy clusters form because the external gravitational influence would diminish the effect of MOND and galaxies cannot form without boosted gravity (whether be it from cold dark matter or modified gravity).

For these reasons, an accurate comparison with the galaxy power spectrum will have to wait until the software of MOND cosmological simulations improves dramatically.

\begin{figure*}
\def\subfigtopskip{0pt} 
\def\subfigbottomskip{4pt}
\def\subfigcapskip{1pt}
\centering
\begin{tabular}{cc}
\subfigure{
\includegraphics[angle=0,width=8.0cm]{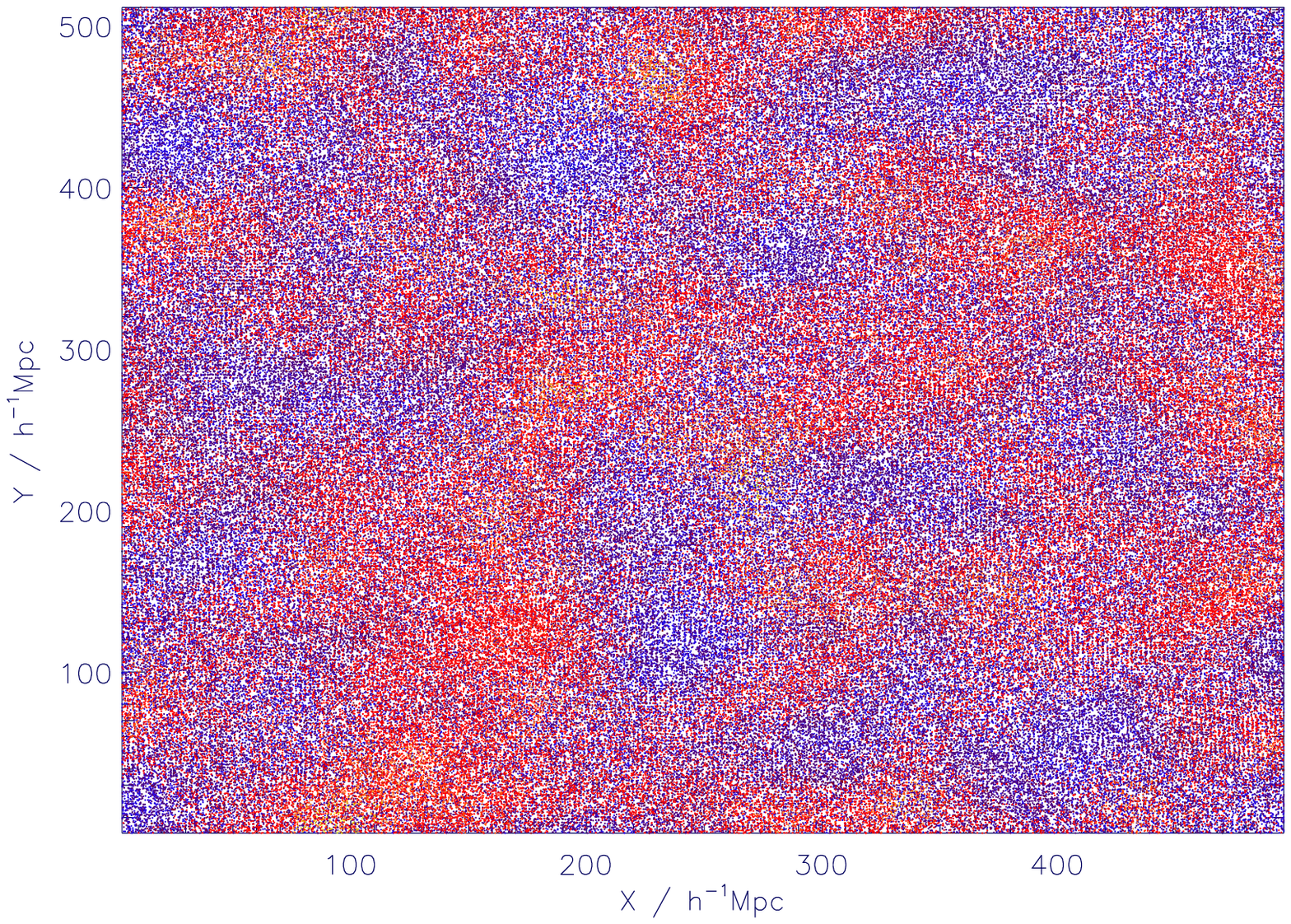}
}
\subfigure{
\includegraphics[angle=0,width=8.0cm]{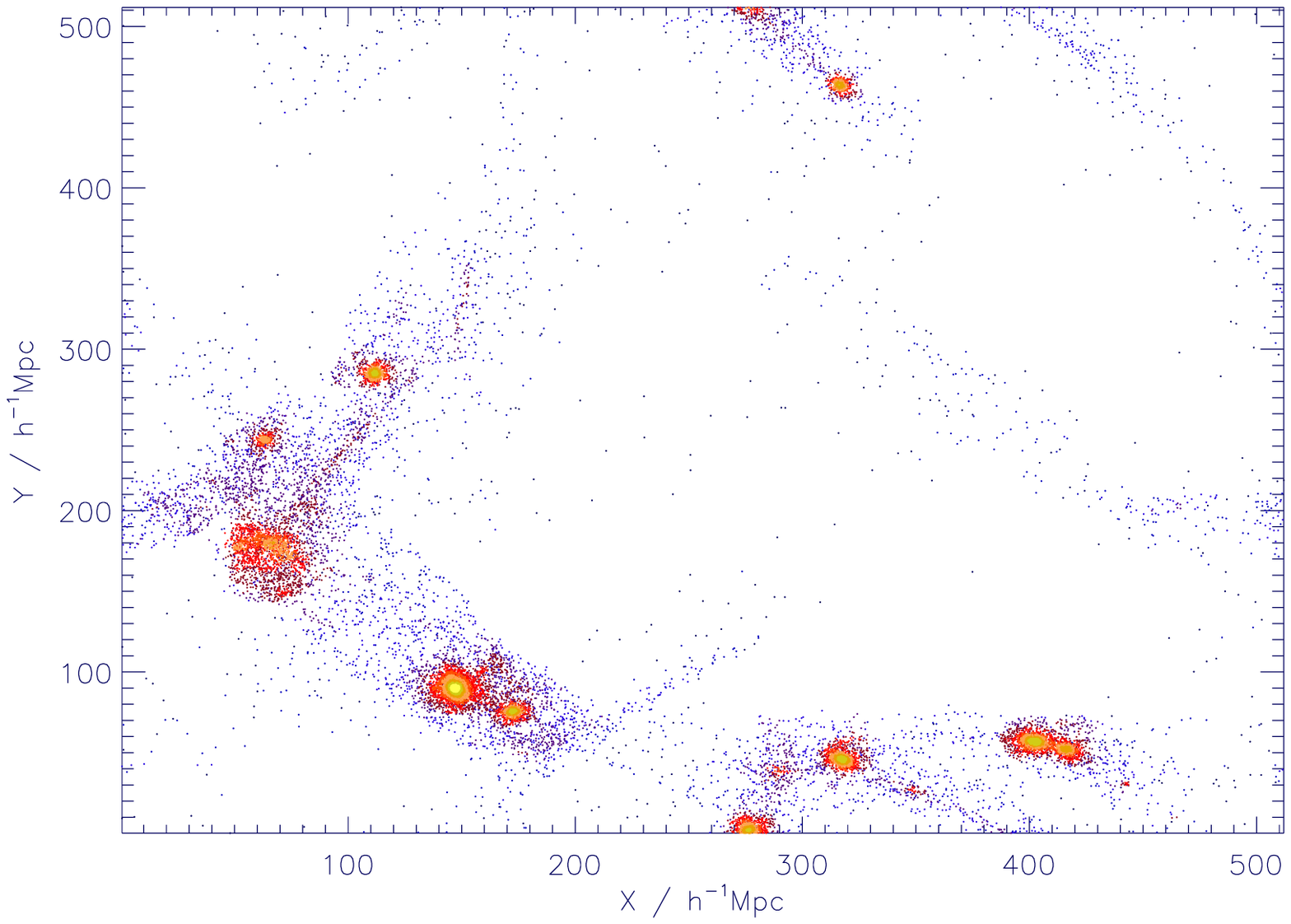}
}
\\
\end{tabular}
\caption{Shows the distribution of particles in a $(512~Mpc/h)\times(512~Mpc/h)\times(50~Mpc/h)$ slice, where the particles are colour coded according to their local 3D density. The left hand panel is a simulation without MOND and the right hand panel is with MOND. The contour levels are very different in the two figures: the left hand panel's extreme density contrast is only 50\% greater than the cosmic mean, whereas the right hand panel's is many thousands of times the cosmic mean.}
\label{fig:2d}
\end{figure*}

 To complement the comparison between structure formation with MOND and without, we also show in Fig~\ref{fig:2d} a two dimensional projection of the density contrast in the two different scenarios. Specifically plotted are the particles in a $50~Mpc/h$ slice along the line of sight, where each particle is given a colour to indicate the local three-dimensional density. The colour contrasts in the two figures are different, the maximum contrast with MOND at this resolution is many thousand times the cosmic mean, without MOND it is barely 50\% . Therefore, any successful cosmology based on sterile neutrinos with an $11~eV$ mass needs a modified gravity theory which boosts perturbations, especially on small scales.

\section{Galaxy clusters}
\subsection{Mass function}
\begin{figure*}
\def\subfigtopskip{0pt} 
\def\subfigbottomskip{4pt}
\def\subfigcapskip{1pt}
\centering
\begin{tabular}{cc}
\subfigure{\label{fig:npro}
\includegraphics[angle=0,width=8.0cm]{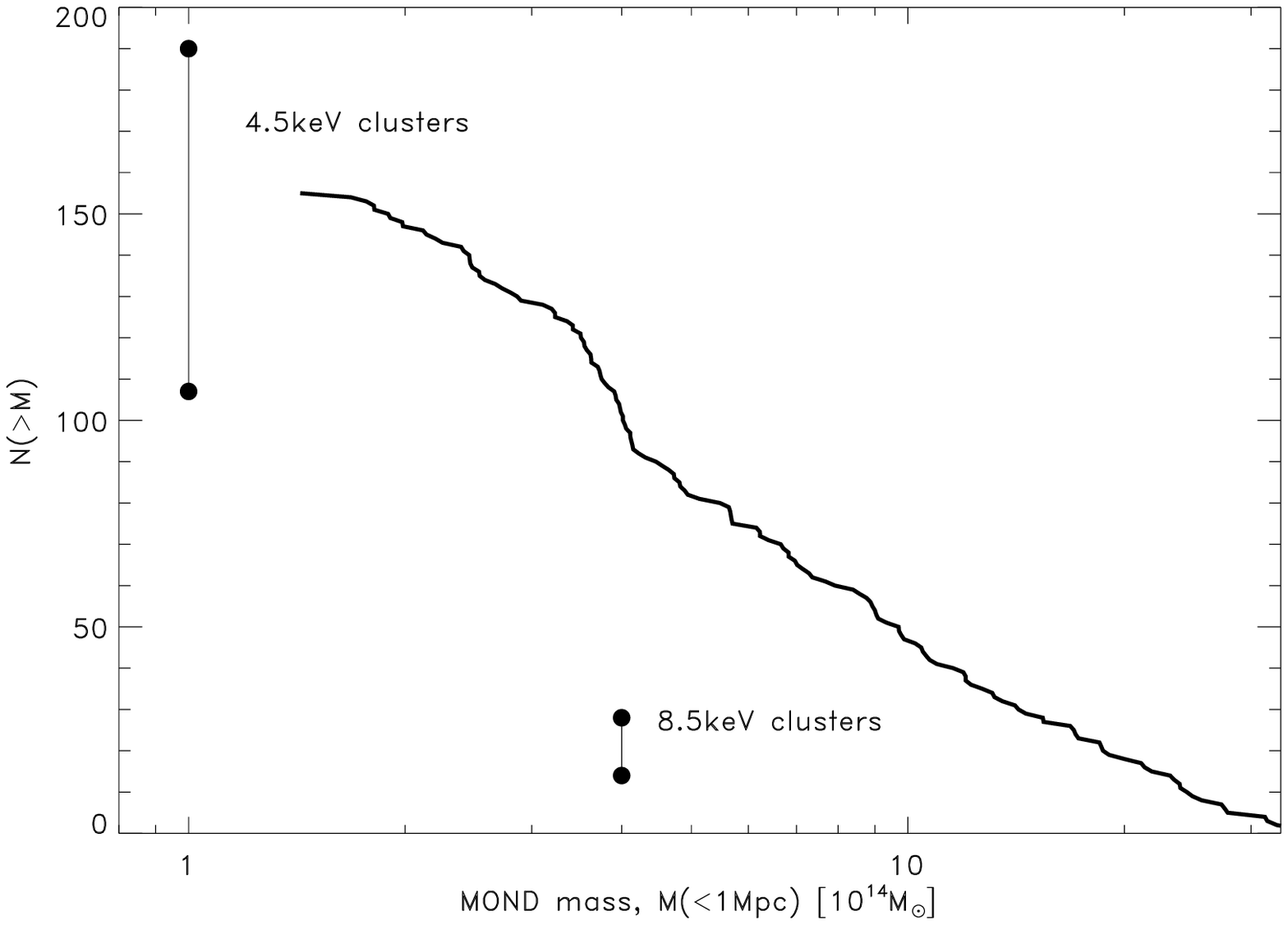}
}
\subfigure{\label{fig:rines}
\includegraphics[angle=0,width=8.0cm]{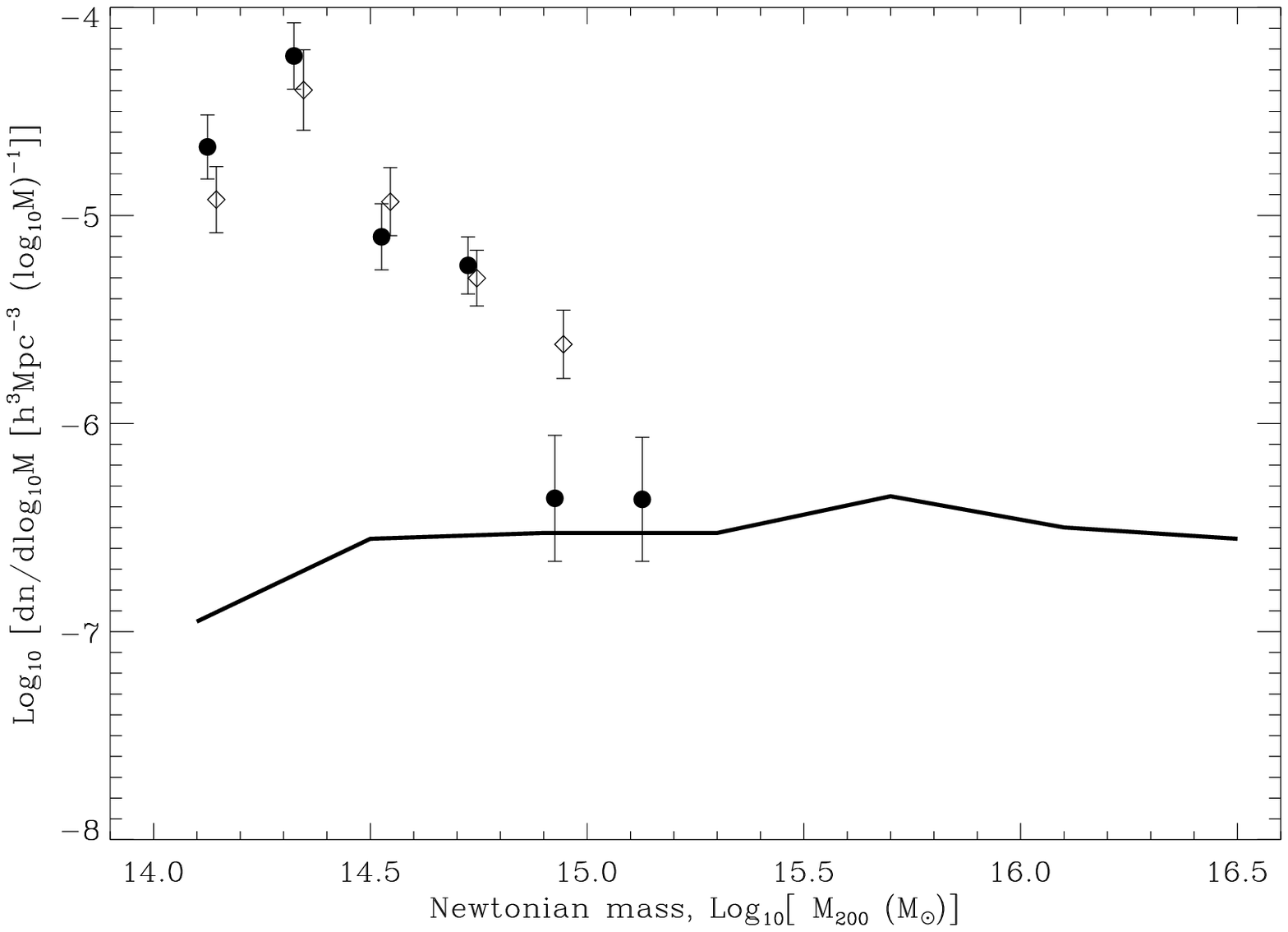}
}\\
\end{tabular}
\caption{(a) In the left hand panel we show the cumulative number of halos in a $(512~Mpc/h)^3$ volume as a function of mass enclosed within $1~Mpc$. Against this curve we have plotted the number of halos hotter than $4.5~keV$ and $8.5~keV$ in such a volume as measured by Pierpaoli, Scott and White (2001). (b) In the right hand panel we plot the number density of halos per $(Mpc/h)^3$ and per logarithmic interval of mass plotted against $log_{10}$ of $M_{200}$. The data points come from Rines, Diaferio \& Natarayan (2008). The filled circles are measurements using a scaling relation between X-ray luminosity and the virial mass, whereas the open diamonds are using the caustic method of Diaferio (1999) to estimate $M_{200}$. The line is using the halos found in our QUMOND simulation.}
\end{figure*}

\protect\label{sec:massfunc}
Despite a comparison with the galaxy power spectrum being unfeasible due to the finite resolution (each particle weighs $7.8\times10^{11}\msun$), it is important that the structures formed in the simulation at $z=0$ resemble the observed distribution and densities of clusters of galaxies - which can be adequately resolved. In \cite{pierpaoli01}, hereafter PSW01, their Fig~6 shows the observed cumulative number distribution of X-ray clusters above a specific temperature. The expected number of X-ray clusters with a temperature larger than $4.5\pm0.5~keV$ and $8.5\pm0.5~keV$ is $107-200$ and $14-28$ respectively in a $(512~Mpc/h)^3$ volume. Upon the $z=0$ output of our simulations, we ran the halo finder of \cite{knebe04} and discarded all halos with less that 1000 particles, so that we could adequately resolve the density profiles of all 155 cluster mass halos. We show in Fig~\ref{fig:npro} the cumulative number of halos in our $(512~Mpc/h)^3$ volume as a function of mass enclosed within $1~Mpc$ ($0.732~Mpc/h$). All these 155 halos had masses within $1~Mpc$ of greater than $10^{14}\msun$. 

In these simulations, we are frequently required to compare the mass distributions of our MONDian halos, with those that a Newtonist would derive. Obviously our halos are composed of baryons and sterile neutrinos and have a specific gravity profile depending on the mass profile - that is, the MONDian mass profile. Similarly, this gravity profile observed by a Newtonist would correspond to another, Newtonian mass profile. In QUMOND, there is a simple algebraic relationship between a spherical MONDian mass profile in MONDian gravity and the Newtonian gravity of that MONDian mass profile, and therefore, the Newtonian mass profile. The Newtonian gravity of the MONDian mass profile $M_m(r)$ is naturally $g_N(r)=GM_m(r)/r^2$ and the MONDian gravity profile of the MONDian mass distribution is $g_M(r)=\nu(g_N(r))g_N(r)$. The final step is to say that this MONDian gravity profile corresponds directly to a (fictitious) Newtonian mass profile i.e. the mass profile a Newtonist would derive from a dynamical probe of some kind, and this is simply $M_N(r)=g_M(r)r^2/G$. Obviously this means the radius at which you compare your mass estimates in the two gravity theories can make a huge difference to your mass ratio, which is worth keeping in mind.

Therefore, at the radius of $1~Mpc$ which we chose to compare with some results from the literature, a MOND mass of $10^{14}\msun$ corresponds to a Newtonian gravity of $g_N(r=1Mpc)=4.42\times10^{-3}\times10^{14}/(10^6)^2=0.442(\kms)^2/pc$. Now since the MOND acceleration constant in these units is $a_o=3.6~(\kms)^2/pc$ we find that $x=0.442/3.6=0.123$, $\nu(y)=3.4$ and therefore the Newtonian mass is $3.4\times10^{14}\msun$. This large a mass is typically found in clusters like A133 and A383 (see \citealt{vikhlinin06}) at 4.1 to 4.8~$keV$ respectively. Similarly, we expect a MOND mass of $4\times10^{14}\msun$ in clusters like the 8-9keV A~2029 and A~2390. For these typically observed MOND masses, we use the range of expected clusters from PSW01 as a gauge of whether our simulation forms the correct number of halos.

It would appear from Fig~\ref{fig:npro} that we form roughly the correct order of magnitude of clusters hotter than $4.5~keV$, but form far too many high mass clusters. This contradicts the naive expectation that galaxy clusters cannot be formed with MOND and $11~eV$ sterile neutrinos. It is in fact reassuring that there is a spectrum of cluster masses formed and that more low mass clusters are formed at the expense of larger clusters and that all the matter does not merely end up in a single cell at the centre of the simulated grid.

We have included a similar test in Fig~\ref{fig:rines} in which is plotted the number density of clusters per $(Mpc/h)^3$ and per interval of $log_{10}M_{200}$. The data points come from \cite{rines08}, hereafter RDN08, where the filled circles are measurements using a scaling relation between X-ray luminosity and the virial mass and the open diamonds are using the caustic method of \cite{diaferio99} (see also \citealt{serra11}) to estimate $M_{200}$. Our estimation using the halos found in our QUMOND simulation is computed in a similar way to the aforementioned comparison with the data of PSW01. Again we begin by initially finding the MONDian mass profile of all the clusters and then convert it to the Newtonian mass which would provide the same gravity as a function of radius. From the Newtonian mass we derived the Newtonian $M_{200}$ which is the enclosed mass at the radius where the density falls to $200$ times the critical density. Given that a Newtonian virial mass of $10^{15}\msun$ at several Mpc would correspond roughly to a MONDian mass of $10^{14}\msun$, our resolution precludes us from comparing with all but that last data point shown in RDN08.

Our model appears to demonstrate the same trend with the RDN08 data as it does with the PSW01 data, except that the RDN08 data probes somewhat smaller cluster masses than the PSW01 sample. By probing cluster temperatures up to $9~keV$, the PSW01 sample is probing Newtonian virial masses up to $2\times10^{15}\msun$ ($log_{10}M_{200}=15.3$), which is marginally larger than the $log_{10}M_{200}=15.1$ limit of RDN08, which by contrast probes far lower virial masses than the PSW01 survey does. This is the reason why our line goes through the last two data points in Fig~\ref{fig:rines} and yet over estimates the number of very massive clusters ($T=8.5~keV$) in Fig~\ref{fig:npro}. Our hypothesis is that if the RDN08 study was extended to $log_{10}M_{200}=15.3$, our current model would overestimate the numbers per mass bin.

Comparing Figs~\ref{fig:npro} and \ref{fig:rines} it is apparent that we have agreement with the moderately massive clusters ($T=4.5~keV$) in Fig~\ref{fig:npro} only because we produce many times more $T>8.5~keV$ clusters than desired. Therefore, Fig~\ref{fig:rines} demonstrates that we severely underpredict the number of $T=4.5~keV$ clusters. This currently is a resolution issue that will be addressed by future simulations. A point worth noting is that the mass function per mass bin as plotted in Fig~\ref{fig:rines} is flat for all halo masses. However, if so much mass was not locked up in these extremely massive MOND halos with $>5\times10^{14}\msun$ within $1~Mpc$ (in fact, there is at least $5\times10^{16}\msun$ locked up in these), then there could be 500 more $10^{14}\msun$ halos produced in principle.

The task now is tuning the available free parameters in order to fit the measured mass function. These include the normalisation of the initial power spectrum, the $\nu$-function as well as two considerably more interesting prospects. The first is how the MOND acceleration constant varies with redshift and the second is how the background expansion of a QUMOND cosmology might vary from the FRW family of models. The first case is almost unconstrained since no detailed studies of the dynamics of galaxies exist at any significant redshift.

Here we make only the hypothesis that the increased energy density of dark energy with increasing scale-factor is linked (perhaps directly) to the increased prevalence of QUMOND fields as the Universe becomes more sparse. If accelerations are much larger than $a_o$ at $z>100$ then there would be no QUMOND fields, but as the Universe expands and accelerations drop below $a_o$ (or whatever value $a_o$ takes at a given redshift) then these QUMOND fields would begin to increase. This is qualitatively the same trend as dark energy takes and is a suggestion that will be expanded upon in a forthcoming paper where we will constrain models of the expansion history of the Universe with supernovae and gamma-ray burst data.

In fact, using a full Bayesian appraoch, \cite{diaferio11} have shown that data from supernovae and gamma-ray bursts can be described by the $\lcdm$ model as well as other more exotic expansion histories. For now it is enough to remark that a different expansion history could be key to the issues mentioned above, with reference to Figs~\ref{fig:npro} and \ref{fig:rines}, for the following reasons: if the formation of the over-abundance of very massive clusters ($> 10^{15}\msun$ in terms of their MOND mass) that we see in our simulations is due to the merging of many subhalos or alternatively due to monolithic collapse, then more rapid expansion at a certain key stage could halt the formation of these superclusters and sustain the abundance of lower mass halos.

 Most likely, the expansion history of the Universe, a varying $a_0$, the $\nu$-function and the normalization of the power spectrum must be traded off against each other. In addition, our resolution limit of roughly $10^{14}\msun$ within $1~Mpc$ must be greatly extended in order to solidify these claims by checking the formation of clusters down to small groups of galaxies, which have been shown in \cite{afb} and \cite{afd} to host dense sterile neutrino halos - $10^{13}\msun$ within $0.2~Mpc$ - and prove to be the strictest test of the Angus09 model.

Although it would appear now to be within reasonable doubt that we could fit the distribution - at least the high temperature end - of massive clusters of galaxies, this in no way gives any information about the formation and distribution of galaxies. As we discussed in \S\ref{sec:ps}, the formation process of galaxies is entirely dependent on perturbations to the distribution of cold gas and the accretion of more gas aided by MONDian gravity. The sterile neutrinos will have no influence except presumably as distant tidal torques.

\subsection{Density profiles}
\begin{figure*}
\def\subfigtopskip{0pt} 
\def\subfigbottomskip{4pt}
\def\subfigcapskip{1pt}
\centering
\begin{tabular}{cc}
\subfigure{\label{fig:dens}
\includegraphics[angle=0,width=8.0cm]{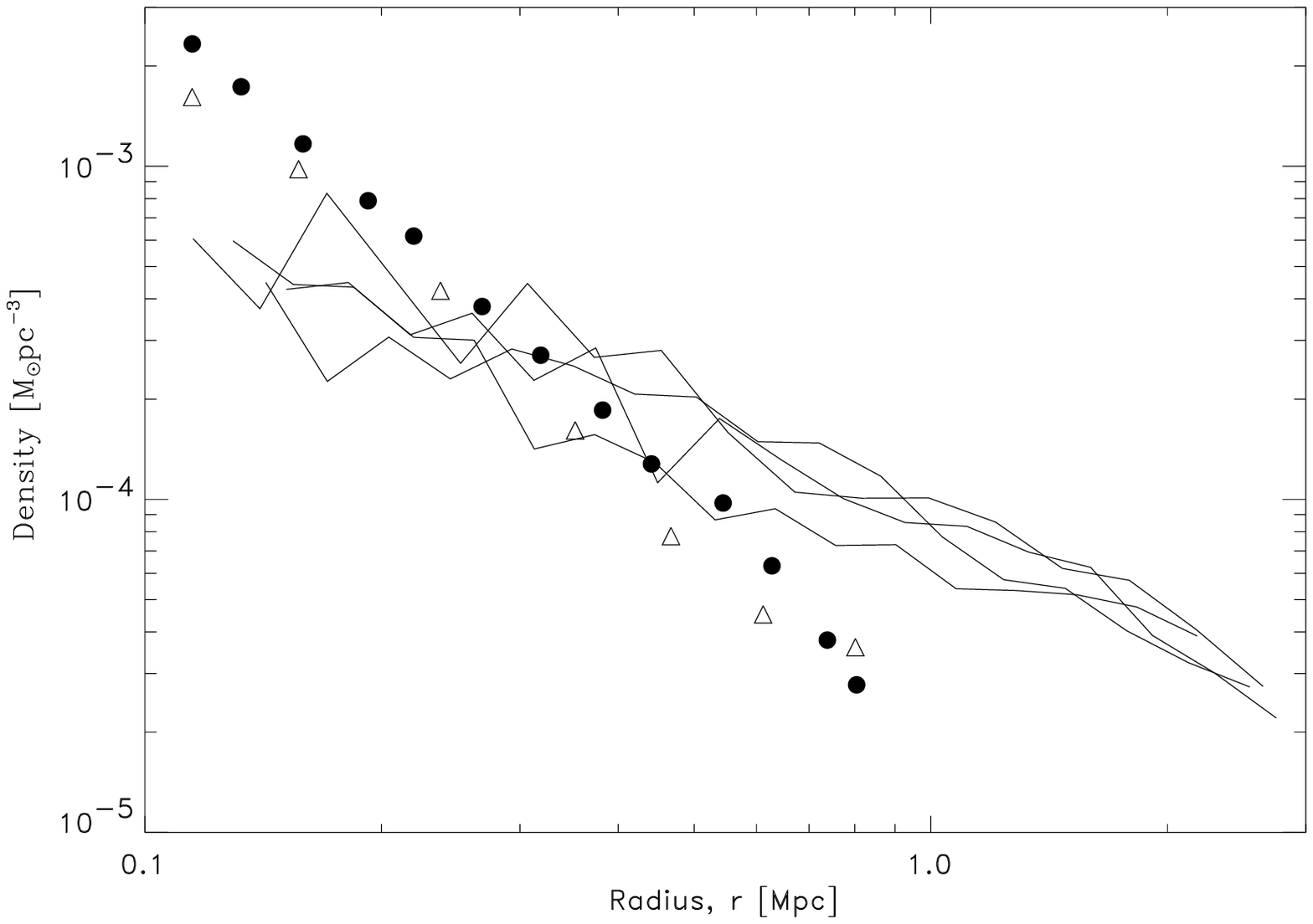}
}

\subfigure{\label{fig:bigclu}
\includegraphics[angle=0,width=8.0cm]{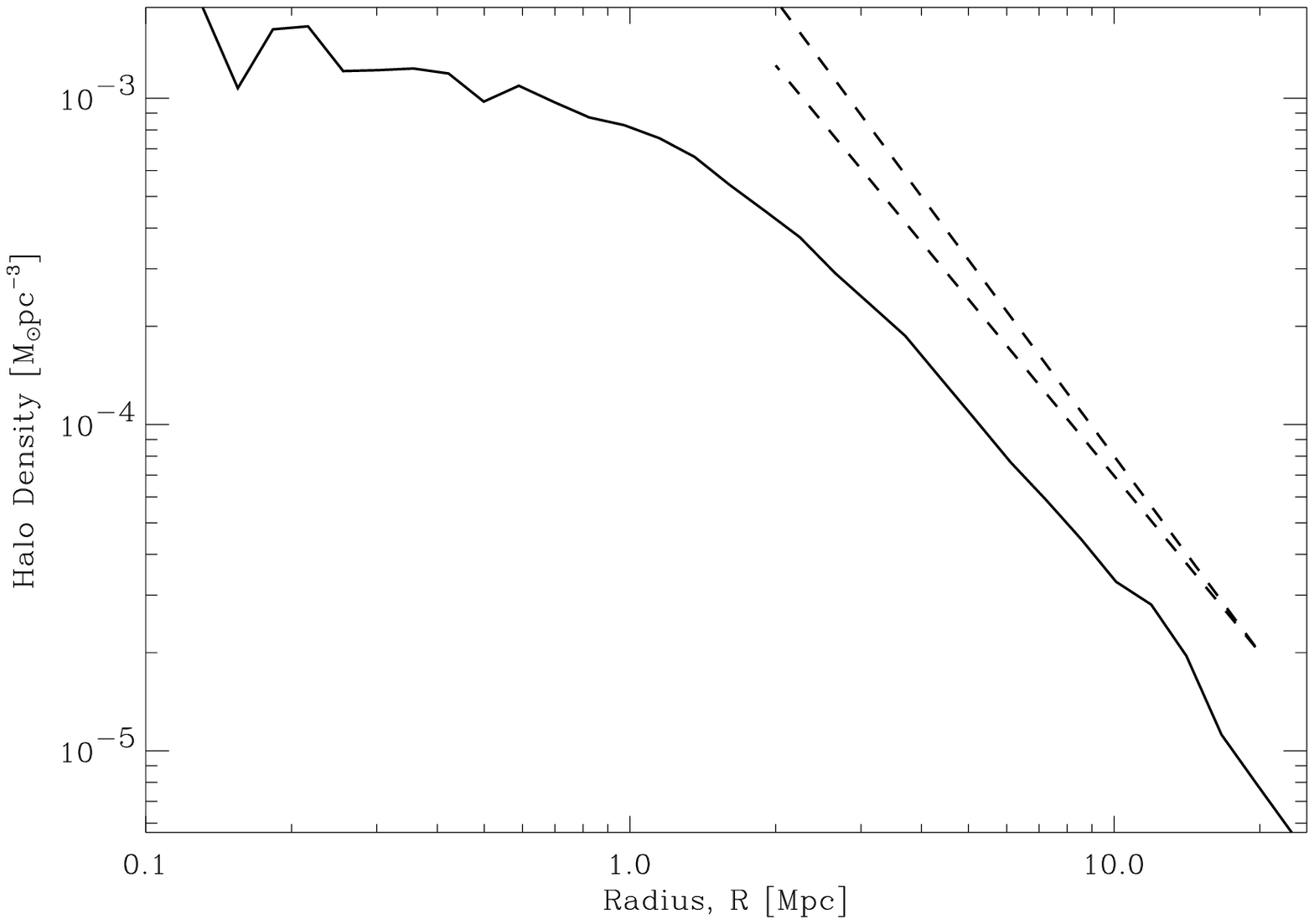}
}

\\
\end{tabular}
\caption{(a) In the left hand panel the lines plotted are the density profiles of cluster sized halos with a similar mass to the rich clusters A~2029 and A~2390, whose MONDian density profiles are also plotted as the triangle and filled circle symbols. We note that the radiis probed here are below the resolution limit of the simulation. (b) In the right hand panel we plot the density profile of the largest cluster we form which extends well into the resolved range and overplotted are logarithmic density slopes of -2 and -1.8 with dashed linetype.}
\end{figure*}

Another thing that is uncertain about galaxy clusters formed from sterile neutrinos under MONDian gravity is their internal structure. In \cite{afd} we extracted the density profiles of the sterile neutrinos required to fit the mass profiles of X-ray clusters measured from the density and temperature of the intracluster medium. After merging the sterile neutrino densities with the small contribution from the X-ray gas, we show in Fig~\ref{fig:dens} that the two largest clusters (A~2029 and A~2390) both had logarithmic density slopes over the range 0.1 to 1.0 Mpc of around ${d\ln\rho \over d\ln r}\approx -2.0$, and certainly greater than -2.5. We plot alongside this the density profiles for four typical halos from our simulation with similar masses. Our clusters have outer density slopes that are significantly shallower than $-2$. The shallower densities are due to a lack of resolution and to avoid the conclusion that this is a generic feature of sterile neutrino halos emerging from cosmological simulations, we also plot in Fig~\ref{fig:bigclu} the density profile of our most massive halo out to $25~Mpc$, where resolution is certainly not a problem. We have overplotted logarithmic slopes of -2 and -1.8, and the latter seems to resemble the outer density profile over an order of magnitude in radius.

In addition to this, baryonic adiabatic contraction in clusters has been shown by \cite{gnedin04} to increase the steepness of the density profiles of cluster dark matter halos in a $\lcdm$ universe, so there remain several reasons to be optimistic about the densities of cluster size halos at intermediate radii. Were we to have perfect resolution, we would expect our halos to resemble the best fits to cluster of galaxy dark matter halos in MOND, at least down to 100~kpc. That is unless there is a scale dependence on cluster halos such that the most massive clusters have significantly shallower, or steeper, outer density profiles.

Related to the discussion in \S\ref{sec:massfunc}, it may be the case that the steepness of the halos is dependent on the redshift of formation which would be different if the MOND acceleration constant was lower at higher redshift. What is certain is that the formation redshift of all clusters will drop if the MOND acceleration constant is lower at high redshift, but the effect on the steepness of the density profiles is less clear. In CDM simulations for example, the concentration parameters of halos are observed to decrease with increasing mass (hence decreasing formation redshift) i.e. smaller halos form earlier and are therefore more dense (see e.g. \citealt{bullock01a}). Presumably this result holds also in MOND, so delaying the formation to lower redshift would decrease the concentration of the sterile neutrino halos. Increasing the amplitude of the initial fluctuations (through the CMB quadrupole) could act in the opposite direction, so perhaps tuning between the two can provide the correct balance between number density of halos and steepness of the density profiles of individual clusters. 


Equally important, the inner profiles of the clusters studied in \cite{afd} were found to reach densities in excess of $10^{-2}\msun pc^{-3}$ at $10~kpc$. This was the case for all but the sub $1~keV$ groups of galaxies. At the moment, our simulations are unreliable below $1000~kpc/h$ and cannot probe such small radii, but it will be vital in future studies to check whether the densities of the sterile neutrino halos can continue to rise all the way to $10~kpc$ and reach the relatively high densities found there in the real cluster halos.

The reason we cannot simply run a simulation with a $16~Mpc/h$ box and immediately find the central densities of the clusters is because that small a box cannot incorporate the influence of the large scale structure on the local environment, which is of crucial importance in MOND cosmological (and ordinary) simulations. Ideally, one should not use a box smaller than we have used here and so a refinement strategy and more than $256^3$ particles will be necessary. In principle, a resimulation technique,
like that used by \cite{springel01}, could be invaluable.

\section{Pairwise velocities}
The relative velocity of the two clusters that comprise the bullet cluster (\citealt{clowe04,clowe06,bradac06}) is a topic of much interest basically because, at first order, it appears to be larger than one could naturally expect in a $\lcdm$ universe (see e.g. \citealt{markevitch02,markevitch07}). \cite{mastro08} have recently shown via hydrodynamic simulations that the infall velocity of the clusters must be $3000~\kms$ within 2-3$R_{200}$ in order to reproduce the observed features of the various shocks. With this in mind, \cite{lee10} have simulated a large cosmological volume ($27~Gpc^3/h^3$) to infer the statistical probability of bullet cluster like events, which they showed to be incredibly low (between $3.3\times10^{-11}$ and $3.6\times10^{-9}$).

\begin{figure}
\def\subfigtopskip{0pt} 
\def\subfigbottomskip{4pt}
\def\subfigcapskip{1pt}
\centering
\begin{tabular}{c}
\subfigure{
\includegraphics[angle=0,width=8.0cm]{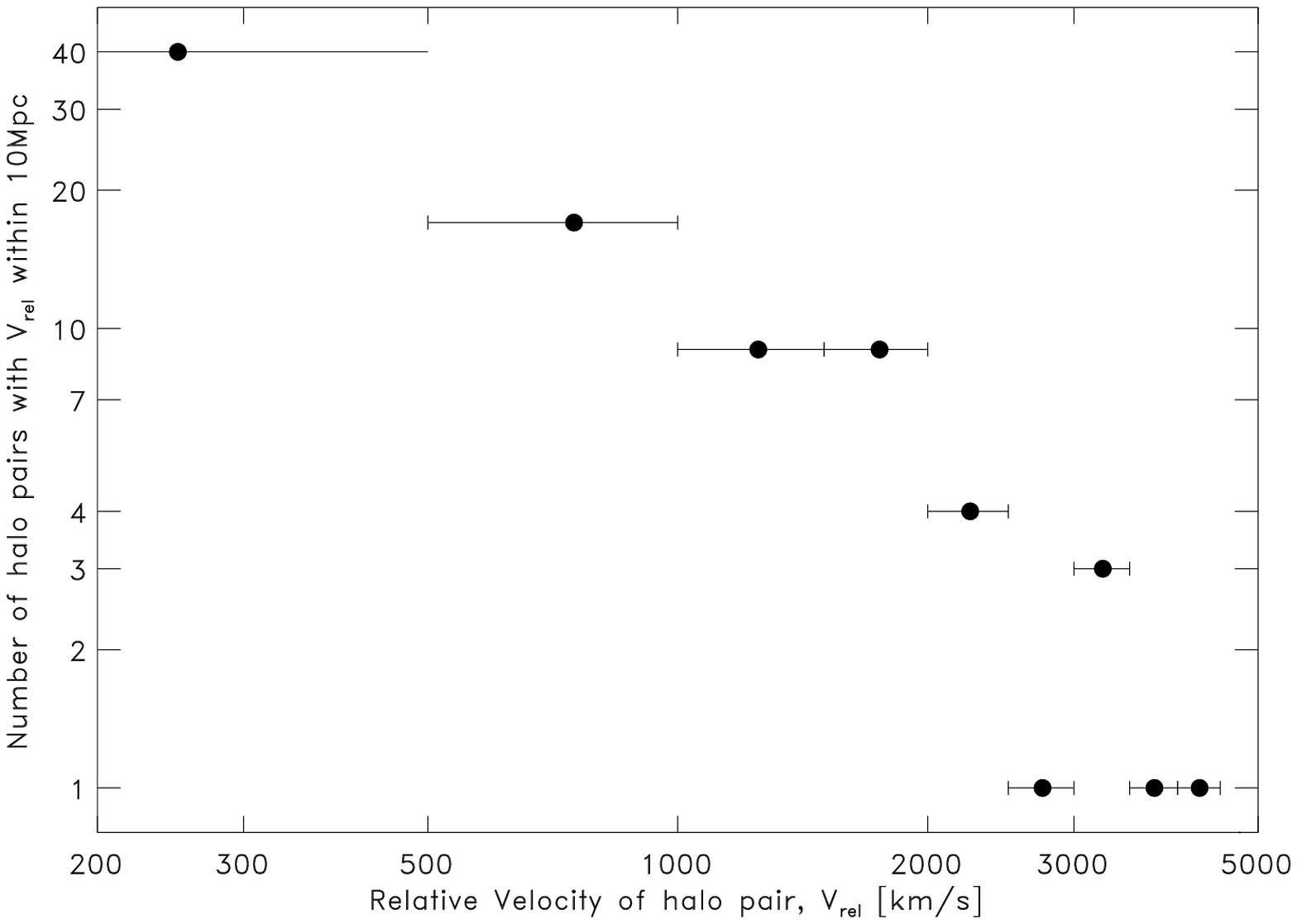}
}\\
\end{tabular}

\caption{Shows the numbers of pairs of halos within a sphere of radius $10~Mpc$ of each other which have a pairwise relative velocity in the range delineated by the horizontal error bars.}
\label{fig:bullet}
\end{figure}

To estimate the likelihood of creating bullet cluster like events in the Angus09 cosmological model, we show the $z=0$ (not $z=0.3$) relative velocities between all pairs of halos within a sphere of $10~Mpc$ from each other, which is in general between 3-6$R_{200}$. We find that out of the $155\times(155-1)/2\approx12000$ potential pairs, there are less than $100$ pairs within $10~Mpc$ of each other and out of these we find $5$ with relative velocities exceeding $3000~\kms$ (see Fig~\ref{fig:bullet}). These velocities will increase with decreasing separation.

This number is inclusive of all bullet cluster sized halos since the sub cluster was observed to have a temperature of $4.5~keV$, which is roughly the limit of our resolution. Although the relative velocities will be lower at $z=0.3$ it still seems plausible that, just as was found by \cite{am08}, MOND has the ability to produce very swift collisions between clusters. With better resolution it might be interesting to check the contrary problem which is if MOND would produce too many bullet cluster like systems.

\section{Conclusion}
Here we introduced a new Particle-Mesh code that accurately solves the modified Poisson equation of Milgrom's QUMOND theory. We presented a set of initial conditions for the Angus09 cosmological model and computed the power spectrum to demonstrate how modified gravity (in particular MOND) is absolutely essential to create large scale structure in a hot dark matter universe.

We ran a simulation from $z=250$ to $z=0$ and identified that we form roughly the correct order of magnitude number of massive clusters of galaxies hotter than $4.5~keV$ and form a reasonable spectrum of masses. Having said that, we considerably underpredict the number of $T=4.5~keV$ clusters and overpredict the number of $T=8.5~keV$ clusters. We suggest that the former problem is a resolution issue that will be dealt with by forthcoming simulations, but that some variation of the available free parameters will be necessary to match the number of very rich clusters of galaxies. The two key variables are how the MOND acceleration constant decreases with increasing redshift and how the expansion history of our model differs from the FRW prediction.

Our resolution was insufficient to compare the density profiles of our halos with those required to fit X-ray clusters in MOND (\citealt{afd}). However, we showed that the logarithmic slope of the largest clusters in our simulation, which extended to beyond $25~Mpc$ and hence well into the resolved range, was compatible with the observed halos. 

To follow up the work of \cite{am08}, which suggested that the large relative velocity of the bullet cluster was a natural result of MOND, we found the numbers of pairs of clusters within $10~Mpc$ of each other and used their relative velocities to compare with the conclusion of \cite{mastro08}. They showed that the relative velocity of a pair needed to be above a certain threshold velocity to generate the observed bow shock and other features of the bullet cluster system. We found that large relative velocities are indeed a natural product of a MOND cosmology.

In the future, it will be vital to amend the code to resolve structures down to below $100~kpc$ and even $10~kpc$. Only then will it be clarified whether or not the cluster halos formed in cosmological N-body simulations with MOND are fully compatible with the observed mass discrepancy in MONDian clusters.

\section{acknowledgments} GWA's research is supported by the National Research Foundation of South Africa and both GWA and AD's research is supported by the University of Torino and Regione Piemonte. Partial support from the INFN grant PD51 and PRIN-MIUR-2008 grant \verb"2008NR3EBK_003" ``Matter-antimatter asymmetry, dark matter and dark energy in the LHC era'' is also gratefully acknowledged. We appreciate comments from Silvio Bonometto, Tom Zlosnik, Martin Feix, Benoit Famaey, Gianfranco Gentile and Kurt van der Heyden.

\end{document}